\documentclass{annurep}
\usepackage{graphicx}
\usepackage{wrapfig}

\newcommand{\beq}{\begin{eqnarray}}
\newcommand{\eeq}{\end{eqnarray}}

\begin{document}

%
\begin{center}
{\large \bf Spin-dipole strengths and tensor correlation effects 
for ${}^{208}{\rm Pb}(p,n)$ at 295 MeV}
\vspace*{0.3cm}

T.~Wakasa for the RCNP-E57 and E59 collaborations\\
{\it Department of Physics, Kyushu University, Higashi,
Fukuoka 812-8581, Japan}
\end{center}

\vspace*{0.5cm}
%
%
 We performed the multipole decomposition analysis (MDA) 
for the ${}^{208}{\rm Pb}(p,n)$ data in order to 
obtain the spin-dipole (SD) strengths separated into 
each $\Delta J^{\pi}$ contribution 
$dB({\rm SD}_{\Delta J^{\pi}};\omega)/d\omega$.
 In the standard MDA \cite{ppnp}, 
the experimentally obtained angular distributions 
$\sigma^{\rm exp}(\theta,\omega)$ of the cross section 
are fitted using the least-squares method using the following 
linear combination of the calculated angular distributions 
$\sigma^{\rm calc}(\theta,\omega)$ for various spin-parity 
transfers $\Delta J^{\pi}$'s:
\begin{equation}
\sigma^{\rm exp}(\theta,\omega) 
=
\sum_{\Delta J^{\pi}}a_{\Delta J^{\pi}}
\sigma^{\rm calc}_{\Delta J^{\pi}}(\theta,\omega),
\end{equation}
where $a_{\Delta J^{\pi}}$ are the fitting coefficients and
the $\sigma^{\rm calc}(\theta,\omega)$ values are obtained 
using distorted wave impulse approximation (DWIA) calculations. 
 In the present MDA,
the polarization observable $D^{\rm calc}(\theta,\omega)$, 
which are sensitive to $\Delta J^{\pi}$ \cite{moss},  
are also evaluated with the DWIA results 
$D^{\rm calc}_{\Delta J^{\pi}}(\theta,\omega)$ for the relevant 
observable by weighting each $\Delta J^{\pi}$ contribution 
$a_{\Delta J^{\pi}}\sigma^{\rm calc}(\theta,\omega)$:
\begin{equation}
D^{\rm calc}(\theta,\omega) 
=
\frac{
\sum_{\Delta J^{\pi}}a_{\Delta J^{\pi}}
\sigma^{\rm calc}_{\Delta J^{\pi}}(\theta,\omega)
D^{\rm calc}_{\Delta J^{\pi}}(\theta,\omega),
}{
\sum_{\Delta J^{\pi}}a_{\Delta J^{\pi}}
\sigma^{\rm calc}_{\Delta J^{\pi}}(\theta,\omega).
}
\end{equation}
 The experimental polarization observables 
$D^{\rm exp}(\theta,\omega)$ are also fitted with 
$D^{\rm calc}_{\Delta J^{\pi}}(\theta,\omega)$.
 Thus the variable $a_{\Delta J^{\pi}}$ are determined 
using the least-squares technique to reproduce 
the cross section and polarization observable data 
simultaneously.

\begin{wrapfigure}{r}{66mm}
  \vspace{-6mm}
  \begin{center}
  \includegraphics[width=66mm]{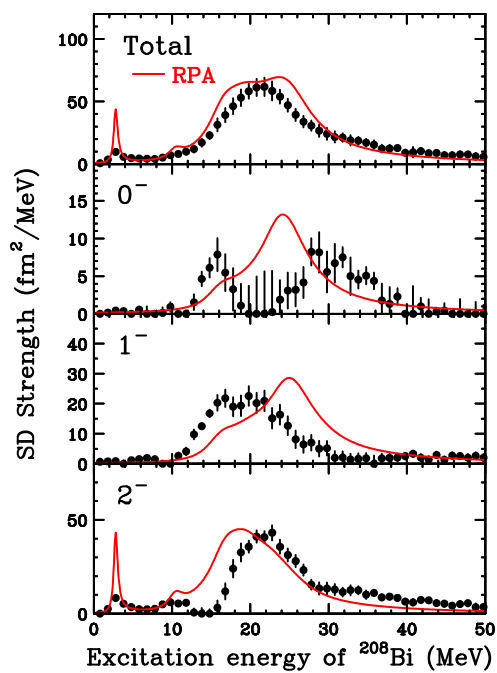}
  \caption{Spin-dipole strength distributions for 
${}^{208}{\rm Pb}(p,n)$.}
  \label{fig:sd}
  \end{center}
\end{wrapfigure}

 The SD strength is obtained by assuming a proportionality 
relation \cite{sdunit}.
 The proportionality relation between 
$dB({\rm SD}_{\Delta J^{\pi}};\omega)/d\omega$
and the relevant cross section, 
$d^2\sigma_{\Delta J^{\pi}}(q,\omega)/d\Omega d\omega$, 
is given by 
$
d^2\sigma_{\Delta J^{\pi}}(q,\omega)/d\Omega d\omega
=
\hat{\sigma}_{{\rm SD};\Delta J^{\pi}}(q,\omega)
dB({\rm SD}_{\Delta J^{\pi}};\omega)/d\omega
$
where 
$\hat{\sigma}_{{\rm SD};\Delta J^{\pi}}(q,\omega)$ is the 
SD cross section per unit SD strength for spin-parity 
transfer $\Delta J^{\pi}$ and depends on both the 
momentum transfer $q$ and the energy transfer $\omega$.
 The $d^2\sigma_{\Delta J^{\pi}}(q,\omega)/d\Omega d\omega$ data 
were taken from the MDA result at 
$4.0^{\circ}$ where the SD transitions are predominantly 
excited.
 The $\hat{\sigma}_{{\rm SD};\Delta J^{\pi}}(q,\omega)$ values 
were obtained using the DWIA calculations.
 Their uncertainties 
have been investigated by evaluating these values 
in DWIA calculations employing 
different optical potential parameters and 
random phase approximation (RPA) 
response functions 
with different Landau-Migdal parameters of 
$g'_{NN}$=0.60$\pm$0.10 and $g'_{NN}$=0.35$\pm$0.16 \cite{ppnp}.
 The uncertainties depend on $\Delta J^{\pi}$, and they are 
about 9\%, 15\%, and 11\% for the 
$0^-$, $1^-$, and $2^-$ transitions, respectively.

 Figure~\ref{fig:sd} shows the preliminary results for the 
SD strength distributions 
obtained in the present analysis.
 The lower three panels are the results of each SD 
strength
$dB({\rm SD}_{\Delta J^{\pi}};\omega)/d\omega$ for
$\Delta J^{\pi}$=$0^-$, $1^-$, and $2^-$, and the top 
panel represents the total SD strength 
by summing up these three strengths.
 The solid curves in Fig.~\ref{fig:sd} are the RPA predictions 
with $g'_{NN}$=0.60 and $g'_{N\Delta}$=0.35 \cite{ppnp}.
 The calculations reproduce the total SD strength reasonably well, 
whereas some discrepancies are found for separated SD strength.
 The centroids of the resonances are slightly lower and higher 
than the theoretical predictions for $1^-$ and $2^-$, respectively.
 These softening and hardening effects observed in 
$1^-$ and $2^-$ distributions 
would be due to the tensor correlation 
effects \cite{sagawa} 
not included in the present RPA calculations.

\end{document}